
\baselineskip=14pt
\magnification=\magstep1
\font\germ=eufm10
\def\hh{\hbox{\germ h}}
\font\bg=cmbx10 scaled 1200
\def\title#1{\hskip1cm{\bg#1} \vskip 5mm}
\def\author#1{\hskip5cm{\bf#1} \vskip 3mm}
\font\fontA=cmbx10 scaled \magstep1

\rightline{hep-th/9307009 \hskip 3mm \ }
\vskip 5truemm
\centerline{\fontA Flat structure for the simple elliptic singularity}
\centerline{\fontA of type $\widetilde {\bf E_6}$ and Jacobi form }
\vskip 5mm

\author{ Ikuo SATAKE }
\font\fontF=cmti8 scaled \magstep0
\centerline{\fontF Department of Mathematics, Faculty of Science, Osaka
University}
\centerline{\fontF Toyonaka-city Osaka 560, Japan}
\medskip

\beginsection \S1. Introduction.
\par\indent
 In order to construct the inverse mapping of the period mapping for the
primitive form for the semi-universal deformation of a simple elliptic
singularity,
K.Saito  introduced in [5] the ``flat structure'' for the extended affine root
system.
In section 3, we construct explicitly the flat theta invariants in the case of
type $E_6$ using the Jacobi form introduced by Wirthm\"uller [7]. Combining the
results of Kato [3], Noumi [4] (explicit description of the flat coordinates ),
this gives an answer to Jacobi's inversion problem (up to linear isomorphism )
of this period mapping for a simple elliptic singularity of type $\tilde E_6$ (
see also [6] ). The details will be published elsewhere.

\beginsection \S2. Jacobi form.
\par\indent
 $\hh_{\bf C}$ is a (complexified) Cartan subalgebra for a fixed simple Lie
algebra of rank $l$.   $\qquad\qquad$
$\hh_{\bf C}^*$:= $Hom_{\bf C}(\hh_{\bf C},{\bf C}).$
$R^{\vee}$: the set of coroots. $W$: Weyl group. $Q(R^{\vee})$: the ${\bf
Z}$-span of $R^{\vee}$. $\qquad\qquad$
$<,>$: the Killing form normalized as $<\alpha,\alpha>=2$  for the highest root
$\alpha$. $\qquad\qquad\qquad$
We identify $\hh_{\bf C}$ with $\hh_{\bf C}^*$ via $<,>$. A symmetric tensor
$$
\tilde I_W:={\partial \over \partial \tau} \otimes {\partial \over \partial
t}+{\partial \over \partial t} \otimes {\partial \over \partial
\tau}+\sum_{i=1}^l {\partial \over \partial z_i} \otimes {\partial \over
\partial z_i},\leqno(2.1)
$$
is defined on ${\bf H}\times \hh_{\bf C}\times {\bf C}\ni (\tau,z,t)$,
where ${\bf H}:=\{\tau \in {\bf C};Im \tau >0\}$, $z_i$ is an orthonormal basis
of $\hh_{\bf C}^*$. The symbol $e(x)$ denotes $exp(2\pi\sqrt{-1}x)$.

\proclaim Definition 2.1.
 A Jacobi form of weight $k$ and index $m$ ( $k,m \in {\bf Z}$ ) is a
holomorphic function
 $\varphi :{\bf H}\times \hh_{\bf C} \times {\bf C} \rightarrow {\bf C}$
satisfying
\item {1)}$\varphi(\tau,z+\lambda+\mu \tau,t-{1\over
2},<\mu,\mu>\tau-<\mu,z>)=\varphi(\tau,z,t)
\hbox{ for any } \lambda,\ \mu \in Q(R^{\vee})$,
\item {2)}$\varphi (\tau,w(z),t)=\varphi (\tau,z,t)\ \hbox{ for any } w\in W$,
\item {3)}$\varphi (\tau,z,t+\alpha)=e(-m\alpha)\varphi(\tau,z,t)\ \hbox{ for
any } \alpha \in {\bf C}$.
\item {4)}$\varphi\left ({a\tau+b \over c\tau+d},{z \over c\tau +d},t+{c<z,z>
\over 2(c\tau +d)}\right )=(c\tau +d)^k \varphi(\tau,z,t)
\hbox{ for any } \left(\matrix{a&b\cr c&d\cr}\right)\in SL(2,{\bf Z})$,
\item {5)}$\varphi$ has a Fourier series expansion of the form
$$
e(-mt)\sum_{n \in {\bf Z}}\phi_n (z) q^n \ \ (\ q=e(\tau))
$$
 with $\phi_n (z)=0 $ if $n<0$.
\par
The vector space of all Jacobi forms of weight $k$ and index $m$ is denoted by
$J_{k,m}$.
Put
$$
J_{**}=\bigoplus _{k,m \in {\bf Z}}J_{k,m},
\ \ \ M_* =\bigoplus _{k \in {\bf Z}}J_{k,0}.\leqno(2.2)
$$
\proclaim Theorem 2.2. ( Wirthm\"uller [7] )
\item{ } $ J_{**} $ is a polynomial algebra over $M_*$, freely generated
by $l+1$ Jacobi forms $\varphi_0,\cdots,\varphi_l$, where $\varphi_i$ is of
weight $k_i$ and index $m_i$ ($m_0 \leq m_1 \leq \cdots \leq m_l$) (except for
the case of type $E_8$).
\par
\proclaim Proposition 2.3.
\item{ } For $\varphi\in J_{k,m}\ and\ \phi\in J_{k',m'}$,
$$
\tilde I_W(d(\eta^{-2k}\varphi),d(\eta^{-2k'}\phi))/\eta^{-2k-2k'} \in
J_{k+k'+2,m+m'}, \leqno(2.3)
$$
where $\eta(\tau):=q^{1/24}\prod_{n=1}^{\infty}(1-q^n) \hbox{ with }q=e(\tau)$.
\par
We take $\varphi=\phi=\varphi_l$.
Then
$$
\psi:=\tilde
I_W(d(\eta^{-2k_l}\varphi_l),d(\eta^{-2k_l}\varphi_l))/\eta^{-2k_l-2k_l}
\leqno(2.4)
$$
is a Jacobi form of weight $2k_l+2$ and index $2m_l$. Since $J_{2,0}=\{0\}$ by
the basic theory of
modular forms, any function multiplied by $\varphi_l^2$ does not appear when
$\psi$ is represented
as a polynomial of $\varphi_0,\cdots,\varphi_l$ over $M_*$.
Also the Jacobi forms $\varphi_0,\cdots,\varphi_l$ generate ``the ring $S^W$''
[5, p.34, (4.3.3)] over $\Gamma({\bf H},{\cal O}_{\bf H})$ [7].
This means that for the ``codimension one case'' ({\it i.e.}\ $m_i<m_l,
i=0,\cdots,l-1$) [5, p.23, (2.4.2)], we can give ``the normalized lowest degree
vector field
$\partial_l$'' defined in [5, p.48, (9.4.1)] by
$$
{\partial \over \partial (\eta^{-2k_l}\varphi_l)}.\leqno(2.5)
$$
By this fact and Prop 2.3, we can express ``the tensor $J^*=\partial_l \tilde
I_W$'' [5, p.49, (9.6.1)] in terms of the Jacobi forms and the Dedekind $\eta$
function.

\beginsection \S3. Jacobi forms and flat theta invariants of type $E_6$
\par\indent

\proclaim Theorem 3.1. ( Wirthm\"uller [7] )
 For the $E_6$ case, the ring $J_{**} $ is a polynomial algebra over $M_*$ on
seven generators
$$
\varphi_0 \in J_{0,1}, \ \ \varphi_1 \in J_{-2,1},\ \ \varphi_2 \in J_{-5,1},\
\  \varphi_3 \in J_{-6,2},\ \  \varphi_4 \in J_{-8,2},\ \ \varphi_5 \in
J_{-9,2},\ \  \varphi_6 \in J_{-12,3}.
$$
\par
 The existence of the above generators is shown by Wirthm\"uller [7].
\proclaim Proposition 3.2.
These generators are unique under the following conditions.
$$
\leqalignno{
\lim_{q\rightarrow 0}\varphi_0=&1/2 e(-t)[S(\omega_1)+S(\omega_6)+90]
,&(3.1)\cr
\lim_{q\rightarrow 0}\varphi_1=&1/2 e(-t)[S(\omega_1)+S(\omega_6)-54]
,&(3.2)\cr
\lim_{q\rightarrow 0}\varphi_2=&e(-t)[S(\omega_1)-S(\omega_6)] ,&(3.3)\cr
\lim_{q\rightarrow 0}\varphi_3=&e(-2t)[-3/2 [S(\omega_3)+ S(\omega_5)]-27
S(\omega_2)+7/8 [S(\omega_1)^2+ S(\omega_6)^2]&(3.4)\cr
&+1/4 S(\omega_1)S(\omega_6)+30 [S(\omega_1)+ S(\omega_6)]-486 ],\cr
\lim_{q\rightarrow 0}\varphi_4=&e(-2t)[-6 [S(\omega_3)+ S(\omega_5)]+36
S(\omega_2)+3/2 [S(\omega_1)^2 + S(\omega_6)^2]&(3.5)\cr
&+S(\omega_1)S(\omega_6)-60 [S(\omega_1)+ S(\omega_6)]+324],\cr
\lim_{q\rightarrow 0}\varphi_5=&e(-2t)[ 6 [S(\omega_3)- S(\omega_5)]-
[S(\omega_1)^2 - S(\omega_6)^2] -42 [S(\omega_1)- S(\omega_6)]] ,&(3.6)\cr
\lim_{q\rightarrow 0}\varphi_6=&e(-3t)[108 S(\omega_4)-324 S(\omega_2)-54
S(\omega_2)[S(\omega_1)+S(\omega_6)]&(3.7)\cr
&-27/2 [S(\omega_1)S(\omega_3)+S(\omega_5)S(\omega_6)]-9/2
[S(\omega_1)S(\omega_5)+S(\omega_3)S(\omega_6)]\cr
&-162 [S(\omega_3)+S(\omega_5)]+15 [S(\omega_1)^3+S(\omega_6)^3]\cr
&-11 S(\omega_1)S(\omega_6)[S(\omega_1)+S(\omega_6)]\cr
&-1620 [S(\omega_1)+S(\omega_6)]-666 [S(\omega_1)^2+S(\omega_6)^2]\cr
&+1836 S(\omega_1)S(\omega_6)-1944],\cr}
$$
where $\omega_k$ is a fundamental weight (see Bourbaki [1]),
$$
S(\omega_k):=\sum_{w\in W/\hbox{ isotropy subgroup of }\omega_k}e(w\cdot
\omega_k),\leqno(3.8)
$$
and $e(\omega_k)$ is a function $e(\omega_k):\hh_{\bf C} \rightarrow {\bf C} $
defined by
$$
\ \ z \mapsto e(<z,\omega_k>).\leqno(3.9)
$$
\par
\proclaim Theorem 3.3.
 Let $\Theta_i$ be the functions satisfying the following relations:
$$
\leqalignno{
&\tilde \varphi_0 =F_1^{'} \eta^2 \Theta_0+F_2^{'} \eta^2 \Theta_1,&(3.10)\cr
&\tilde \varphi_1 =F_1 \eta^2 \Theta_0+F_2 \eta^2 \Theta_1,&(3.11)\cr
&\tilde \varphi_2 =\eta^2 \Theta_2,&(3.12)\cr
&\tilde \varphi_3 =F_3^{'} \eta^2 \Theta_3+F_4^{'} \eta^2 \Theta_4-{3 \over
2\cdot 12^3}(\tilde G_3 \tilde \varphi_0^2-2\tilde G_2^2 \tilde \varphi_0
\tilde \varphi_1+\tilde G_2 \tilde G_3 \tilde \varphi_1^2),&(3.13)\cr
&\tilde \varphi_4 =F_3 \eta^2 \Theta_3+F_4 \eta^2 \Theta_4+{2 \over
12^3}(\tilde G_2 \tilde \varphi_0^2-2\tilde G_3 \tilde \varphi_0 \tilde
\varphi_1+\tilde G_2^2 \tilde \varphi_1^2),&(3.14)\cr
&\tilde \varphi_5 =-2/3  \eta^2 \Theta_5,&(3.15)\cr
&\tilde \varphi_6 ={-6 \over 2\pi \sqrt{-1}}\vartheta_6 &(3.16)\cr
+{1 \over 12^3}&\left[-6 \tilde G_3 \tilde \varphi_0 \tilde \varphi_3+3/2
\tilde G_2^2 \tilde \varphi_0 \tilde \varphi_4+6 \tilde G_2^2 \tilde \varphi_1
\tilde \varphi_3-3/2 \tilde G_2 \tilde G_3 \tilde \varphi_1 \tilde \varphi_4
\right]\cr
+{1 \over 12^6}&\left[-(5\tilde G_2^3+7\tilde G_3^2) \tilde \varphi_0^3
+36 \tilde G_2^2 \tilde G_3 \tilde \varphi_0^2 \tilde \varphi_1
-3 \tilde G_2 (5\tilde G_2^3+7\tilde G_3^2) \tilde \varphi_0 \tilde \varphi_1^2
+\tilde G_3 (8\tilde G_2^3+4\tilde G_3^2) \tilde \varphi_1^3\right],\cr}
$$
where
$$
\leqalignno{
\tilde \varphi:=&\eta^{-2k_i} \varphi \hbox{ for }\varphi \in
J_{k_i,m_i},&(3.17)\cr
G_2:=&(4\pi^4/3)^{-1}g_2(\tau)=(4\pi^4/3)^{-1}60\sum_{m,n\in {\bf Z},(m,n)\not=
(0,0)}{1\over (m\tau +n)^4}\in J_{4,0},&(3.18)\cr
G_3:=&(8\pi^6/27)^{-1}g_3(\tau)=(8\pi^6/27)^{-1}140\sum_{m,n\in {\bf
Z},(m,n)\not= (0,0)}{1\over (m\tau +n)^6}\in J_{6,0},&(3.19)\cr
\vartheta_6:=&\Theta_6-2{\eta^{'} \over
\eta}[\Theta_0\Theta_3+\Theta_1\Theta_4+\Theta_2\Theta_5],&(3.20)\cr
F_i^{'}:=&{-2 \over \pi \sqrt{-1}}\eta^{-4}{dF_i \over d\tau}\ (i=1,2),\
F_i^{'}:={-3 \over 2\pi \sqrt{-1}}\eta^{-4}{dF_i \over d\tau}\
(i=3,4),&(3.21)\cr
F_i:=&f_i(z(\tau))\ \ (z(\tau):={1\over2}\left[{\sqrt{27} \over
\sqrt{-1}}{g_3(\tau) \over (2\pi)^6\eta(\tau)^{12}}+1\right]),&(3.22)\cr}
$$
$f_1(z)\hbox{ and } f_2(z)$ are solutions of Gauss' hypergeometric equation:
$$
z(1-z){d^2f(z) \over dz^2}+({2 \over 3}-{4 \over 3}z){df(z) \over
dz}+{5\over12}f(z)=0 ,\leqno(3.23)
$$
$f_3(z)\hbox{ and } f_4(z)$ are solutions of Gauss' hypergeometric equation:
$$
z(1-z){d^2f(z) \over dz^2}+({2 \over 3}-{4 \over 3}z){df(z) \over
dz}+{1\over12}f(z)=0 ,\leqno(3.24)
$$
satisfying
$$
\left[\matrix{F_1^{'}&F_2^{'}\cr F_1&F_2\cr}\right]
\left[\matrix{F_3^{'}&F_3\cr F_4^{'}&F_4\cr}\right]
=\left[\matrix{{1 \over 12}\tilde G_2^2&{1 \over 3}\tilde G_3\cr {1 \over
12}\tilde G_3&{1 \over 3}\tilde G_2\cr}\right]. \leqno(3.25)
$$
Then the functions $\Theta_i $ are flat theta invariants [5] and satisfy the
following identity:
$$
{2 \over (-2\pi \sqrt{-1})^2}{\partial \over \partial \tilde \varphi_6}\tilde
I_W(d\Theta_i,d\Theta_j)\leqno(3.26)
$$
$$
=\left(
\matrix{
0&0&0&0&0&0&0&1\cr
0&0&0&0&1&0&0&0\cr
0&0&0&0&0&1&0&0\cr
0&0&0&0&0&0&1&0\cr
0&1&0&0&0&0&0&0\cr
0&0&1&0&0&0&0&0\cr
0&0&0&1&0&0&0&0\cr
1&0&0&0&0&0&0&0\cr}
\right)
$$
$$
\ \ ( i,j=-1,0,\cdots,6,\ \Theta_{-1}:=\tau\ ).
$$
\par
\par\noindent
In the proof of Theorem 3.3, we use the following fact:
$$
\eta^{-4}{2 \over (-2\pi \sqrt{-1})^2}{\partial \over \partial \tilde
\varphi_6}\tilde I_W(d\tilde \varphi_i,d\tilde \varphi_j)\leqno(3.27)
$$
$$
=\left(
\matrix{
0&0&0&0&0&0&0&{-6 \eta^{-4}\over 2\pi \sqrt{-1}} \cr
0&0&0&0&{1 \over 12} \tilde G_2^2& {1 \over 3} \tilde G_3 &0&3\tilde G_2 \tilde
\varphi_1\cr
0&0&0&0&{1 \over 12} \tilde G_3 &{1 \over 3} \tilde G_2 &0&2\tilde \varphi_0\cr
0&0&0&0&0&0&-{2 \over 3}&0\cr
0&{1 \over 12} \tilde G_2^2& {1 \over 12} \tilde G_3 &0&{1 \over 2} \tilde G_2
\tilde \varphi_1&{4 \over 3} \tilde \varphi_0&0&4\tilde \varphi_1^2+{1 \over 2}
\tilde G_2 \tilde \varphi_4 \cr
0&{1 \over 3} \tilde G_3 &{1 \over 3} \tilde G_2 &0&{4 \over 3} \tilde
\varphi_0&{8 \over 3} \tilde \varphi_1&0&4\tilde \varphi_3\cr
0&0&0&-{2 \over 3}&0&0&0&0\cr
{-6 \eta^{-4}\over 2\pi \sqrt{-1}}&3\tilde G_2 \tilde \varphi_1&2\tilde
\varphi_0&0&4\tilde \varphi_1^2+{1 \over 2} \tilde G_2 \tilde \varphi_4&4\tilde
\varphi_3&0&* \cr}
\right)
$$
$$
\ \ ( i,j=-1,0,\cdots,6,\ \tilde \varphi_{-1}:=\tau\ ).
$$
\proclaim Remark.
{}From the above results, we obtain that
$$
{\bf C}\Theta_0 \oplus {\bf C}\Theta_1 \oplus {\bf C}\Theta_2 ={\bf C}
\eta^{-2} \chi_{\Lambda_0} \oplus {\bf C} \eta^{-2}\chi_{\Lambda_1} \oplus {\bf
C} \eta^{-2}\chi_{\Lambda_6},
$$
where $\Lambda_0,\ \Lambda_1\hbox{ and } \Lambda_6$ are fundamental weights of
level 1 of $E_6^{(1)}$ type affine Lie algebra, $\chi_{\Lambda}$ is a
normalized character of $E_6^{(1)}$ type ( see [2, p.226] ).
\par

\proclaim References.
\par
\item{[1]} Bourbaki, N., Groupes et alg\`ebres de Lie, Chapitres 4,5 et 6,
Paris, Hermann, 1969.
\item{[2]} Kac, V.G., Infinite dimensional Lie algebras, Third edition, New
York, Cambridge Univ. Press, 1990.
\item{[3]} Kato, M., The flat coordinates of universal unfoldings of $\tilde
E_6, \tilde E_7$, Bull. Coll. Sci., Univ. Ryukyu, 42 (1986), 5-10.
\item{[4]} Noumi, M., Flat coordinate system of type E, $\tilde E_6, \tilde
E_7, \tilde E_8$, Tables (1986), unpublished.
\item{[5]} Saito, K., Extended affine root system II, Publ.RIMS, 26(1990),
15-78.
\item{[6]} Saito, K., Period mapping associated to a primitive form, Publ.
RIMS, 19 (1983),1231-1264.
\item{[7]} Wirthm\"uller, K., Root systems and Jacobi forms, preprint(1989).

\end